# Mechanism of spin crossover in LaCoO$_3$ resolved by shape magnetostriction in pulsed magnetic fields

M. Rotter[1], Z. -S. Wang[2, *], A. T. Boothroyd[3], D. Prabhakaran[3], A. Tanaka[4], and M. Doerr[5]

[1]Max Planck Institute for Chemical Physics of Solids, Nöthnitzerstr. 40, D-01187 Dresden, Germany

[2]Hochfeld-Magnetlabor Dresden, Helmholtz-Zentrum Dresden-Rossendorf, D-01314 Dresden, Germany

[3]Department of Physics, Clarendon Laboratory, University of Oxford, Parks Road, Oxford OX1 3PU, UK

[4]ADSM, Hiroshima University, Higashi-Hiroshima 739-8530, Japan

[5]Institut für Festkörperphysik, Technische Universität Dresden, D-01162 Dresden, Germany

* z.wang@hzdr.de

In the scientific description of unconventional transport properties of oxides (spin-dependent transport, superconductivity etc.), the spin-state degree of freedom plays a fundamental role. Because of this, temperature- or magnetic field-induced spin-state transitions are in the focus of solid-state physics. Cobaltites, e.g. $LaCoO_3$, are prominent examples showing these spin transitions. However, the microscopic nature of the spontaneous spin crossover in $LaCoO_3$ is still controversial. Here we report magnetostriction measurements on $LaCoO_3$ in magnetic fields up to 70 T to study the sharp, field-induced transition at $H_c \approx 60$ T. Measurements of both longitudinal and transversal magnetostriction allow us to separate magnetovolume and magnetodistortive changes. We find a large increase in volume, but only a very small increase in tetragonal distortion at $H_c$. The results, supported by electronic energy calculations by the configuration interaction cluster method, provide compelling evidence that above $H_c$ $LaCoO_3$ adopts a correlated low spin/high spin state.

Perovskite cobaltites possess the so-called spin state degree of freedom, i.e. the existence in octahedrally-coordinated $Co^{3+}$ of energetically proximate low spin (LS, S=0), high spin (HS, S=2), and intermediate spin (S=1, IS) electronic configurations, and the possibility to transition between these states with increasing temperature or magnetic field[1, 2]. This spin state degree of freedom is generally considered to play a key role for the unconventional transport properties of the cobaltites. $LaCoO_3$ is well known for exhibiting an unconventional temperature-induced spin state transition of the $Co^{3+}$ ions from LS to HS[3]. However, the explanation of the spin crossover in $LaCoO_3$ is a matter of ongoing scientific debate. The existence of an intermediate spin state (IS, $S$=1)[4,5] as well as fast oscillation between the LS and HS spin states are discussed as possible scenarios. Several experimental techniques have been employed in order to clarify the mechanism, for example X-ray absorption, magnetic dichroism and electron spin resonance spectra were reproduced successfully by a configuration interaction cluster calculation based on a lattice-aided thermal population of the HS state[3]. A corresponding anomalous volume increase at $LaCoO_3$ has been observed in the thermal expansion[4, 6]. The anomaly can be suppressed by pressure[7]. However, temperature and pressure are isotropic perturbations of the systems and cannot establish an unambiguous interpretation of the microscopic nature of the spin crossover.

Recently, a sharp transition has been found in magnetisation and longitudinal magnetostriction experiments on $LaCoO_3$ in pulsed magnetic fields (60 T)[5,8,9]. Two possible models for this transition are (i) a LS to IS level crossing, and (ii) the formation of a superlattice of LS and HS sites. In model (i), a Jahn-Teller type of lattice distortion can be expected, but the volume change at the transition should be small, whereas in model (ii), there will be a large volume change associated with the formation of HS $Co^{3+}$ but almost no John-Teller distortion. Note that local probes such as EXAFS were not able to find any static or dynamic Jahn Teller distortions induced by temperature or doping[10] which favours the LS-HS model (ii). But, though important and necessary for an ultimate clarification, no magnetoelastic investigations could be done at the spin-crossover field of about 60 T. This lack can be closed by this paper.

Here we show that for the spin crossover in high magnetic field model (i) and (ii) can be distinguished by combining longitudinal and transverse magnetostriction results to isolate the change in volume and distortion at the transition. Our results provide compelling evidence in favour of model (ii) and are one of the strongest arguments to come to this final conclusion.

**Results**

**Magnetostriction measurements**. In order to access the spin transition field[5], the magnetostriction experiments were performed in high magnetic fields up to 70 T on a single crystal of $LaCoO_3$. The optical Fibre Bragg grating (FBG) method[11] is the best suitable procedure for striction experiments in pulsed magnetic fields with millisecond pulse duration. It was successfully tested in longitudinal geometry up to nearly 100 T[8]. The experimental innovation of the present work is to make use of the optical FBG method[11] in pulsed fields up to 70 T in order to do transversal magnetostriction measurements. The combination of longitudinal and transversal measurements allow us to calculate the volume or distortion of $LaCoO_3$ in field, by addition or subtraction of the data curves, respectively (see Fig. 2).

Figure 1a presents the isothermal longitudinal magnetostriction $\varepsilon_l$ measured at three temperatures (1.5 K, 24 K and 42 K) in order to follow the evolution of phase transitions. Starting the magnetostriction FBG measurements from the non-magnetic LS ground state of the $Co^{3+}$ ion in zero field, we detected the first spin crossover transition at $H_c = 62$ T at $T = 1.5$ K, in agreement with ref. 5. The transition is hysteretic (1st order phase transition), i.e. the transition back into the low-field phase happens only at 54 T in decreasing field. This hysteresis becomes narrower at higher temperatures, for example the transition fields are 60 T and 58 T at 24 K. The curve shape in both cases is characterized by a sharp expansion followed by a further increase of the sample length and saturation at about 70 T. In contrast, the curve at 42 K shows a continuous increase of the sample length, and no phase transition is observed within the available field range. Measurements in transversal geometry at similar temperatures, shown in Fig. 1b, confirm the values of the transition fields. The transversal magnetostriction $\varepsilon_t$ shows initially a sharp expansion, followed by a contraction in higher fields. The results in Fig. 1 are the key experimental findings of our study.

From both the longitudinal and transversal striction components, the volume effect $\varepsilon_v = \varepsilon_l + 2\varepsilon_t$ and the tetragonal distortion $\varepsilon_d = \varepsilon_l - \varepsilon_t$ were calculated, see Figs. 2a and 2b. The analysis reveals a large volume jump of the order of $6 \times 10^{-3}$, but a very small distortion at the field-induced spin state transition. The absolute value of the calculated distortion is zero up to the transition field and then increases to about $1 \times 10^{-3}$ at 70 T. In contrast to the volume jump (a first order transition), the distortion increases smoothly for $H > H_c$, which is characteristic of a second-order effect. This behaviour hints to an absence of Jahn-Teller distortion. Although a spatially alternating distortion could result in a similar vanishing of a general distortion, this scenario would cause domains and crystal breaks and, therfore, is not likely.

Although a bulk probe such as magnetostriction is not able to determine the spin state of $Co^{3+}$ unambiguously, the large volume magnetostriction and virtual absence of a Jahn-Teller distortion at the spin crossover already points towards model (ii), the stabilization of a LS-HS superlattice. The realistic model calculations described next put this outcome on a solid quantitative footing.

**Theoretical analysis.** Our theoretical analysis is based on the configuration interaction cluster method, which takes into account the full multiplet structure of $Co^{3+}$. The XTLS 9.0[12] code was used. These calculations are able to reproduce many properties of $LaCoO_3$, for example the spectroscopic results of ref. 3, and are similar to a quantum chemical approach[13,14]. The cluster parameters corresponding to the LS ground state of $LaCoO_3$ have been taken from ref. 3. The effect of distorting the $Co-O_6$ octahedron has been modeled by combining a point charge model for the crystal field and Harrison's rules for the hybridisation matrix elements. We modified the ionic crystal field parameters and the hybridisation parameters with respect to ref. 3 as follows: the charge of a nearest-neighbor point charge model for the ionic crystal field was chosen such that for zero distortion the value of $10Dq$ as given in ref. 3 is obtained. The ionic crystal field of the distorted octahedron was evaluated using this point charge model. Similar, the hybridisation parameters of the Co $d$-electrons and oxygen ligand $p$-electrons were estimated according to the distance-to-the-power-7/2 law by Harrison[15], starting from $V_{pd\sigma}$ = -1.7 eV for a Co-O distance of 1.925 Å[3]. We would like to point out that the results of the calculations, which we discuss below, do not depend strongly on the specific values of the parameters which we have chosen. Taking other similar and reasonable values will lead to the same conclusions.

Using this approach we obtained the energies of the different spin states of the $Co^{3+}$ ion as a function of volume strain (Fig. 3) and tetragonal lattice distortion (Fig. 4). The results of the model calculation show that a small increase in volume $\varepsilon_v$ = 0.02 will stabilize a HS state (Fig. 3), whereas the IS spin state cannot be stabilized by a volume strain. A tetragonal distortion is needed to stabilize the IS. From Fig. 4 we see that for $\varepsilon_d$ = -0.13 to -0.04 a HS state is stable, and for displacements larger than ±0.13 an IS state would become stable. We conclude from these calculations that a huge tetragonal distortion would be required for a $Co^{3+}$ octahedron to transform to the IS state.

In comparing the model calculations to our experimental results we must consider in addition to the electronic energies of the Co $d$ and O $p$ states (shown in Figs. 3 and 4) also the elastic energy of the lattice. For example, if one Co-$O_6$ octahedron goes to the HS state by a volume expansion according to Fig. 3, the neighboring octahedron will experience a decrease of their volume stabilizing their LS state. Such a volume expansion of one octahedron will also cost elastic energy and thus will not happen at zero magnetic field and low temperature.

In order to interpret the behavior in magnetic field, the Zeeman term due to the applied magnetic field and the exchange energy has to be taken into account. These terms will become important if some of the ions undergo a spin state transition from the LS ground state into an IS or HS state. Both energy terms may lower the total energy and thus a transition to a phase with several $Co^{3+}$ ions in IS and HS states becomes possible.

Before going into further details we would like to stress that a Zeeman splitting of the HS (S=2) state without lattice relaxation would lead to a transition field much larger than 60 T. The energy lowering of the HS state caused by the Zeeman splitting at 60 Tesla is around $2\mu_B HS \sim 0.014$ eV. Note that the low-field susceptibility and the temperature dependence of zero-field HS to LS population ratio[3] can be explained only if the energy difference between LS and HS state of the Co-$O_6$ octahedron without lattice relaxation, i.e., no expansion, is more than 0.05 eV (see Fig. 3). Therefore the energy splitting of $2\mu_B HS \sim 0.014$ eV caused by the Zeeman term alone at 60 T cannot induce a LS-HS transition. The HS (S=2) state has definitely to be connected with an expanded Co-$O_6$ octahedron.

We turn now to a quantitative comparison of the results of our model calculation with the experimental data for LaCoO$_3$. Because not all Co$^{3+}$ ions contribute in the transition an interpretation of magnetostriction also needs to consider the magnetisation[5,9], which increases to about 0.5 $\mu_B$ / Co at $H_c$. The two simple ways in which this could be interpreted are:

(i) in an IS scenario, a fraction of 25% of Co IS ($S$=1) ions each carry a moment of 2 $\mu_B$. These drive the magnetostriction transition, while the other 75% of Co ions remain in the LS ground state at $H_c$.

(ii) in a HS scenario, a fraction of 12.5% of Co HS ($S$=2) ions each carrying a moment of 4 $\mu_B$. The remaining 87.5% of Co remain in the LS state at the transition.

In scenario (i), the IS can only be stabilised if each IS Co-O$_6$ octahedron is distorted by at least $\varepsilon_d^{one-oct} \approx 0.13$ (see Fig. 4). Assuming that only 25% are in the IS state this would give an average magnetostriction for the whole crystal of $\varepsilon_d^{(i)} \approx 0.13/4 = 0.033$. This expectation has to be contrasted with the experimentally-measured distortion at the transition field $H_c$ which is only $\varepsilon_d \approx 0.002$ (see Fig. 2). We emphasize, that also spatially varying local distortions of Co-O octahedrons resulting macroscopically in a small $\varepsilon_d$ cannot be reconciled with the anisotropic nature of the magnetic field, which is used to induce the spin crossover. Therefore, we conclude that scenario (i) cannot be used to interpret our experimental data

We turn now to model (ii). In this scenario a Co$^{3+}$ HS octahedron can be stabilized by a volume increase of $\varepsilon_v^{one-oct} \approx 0.024$ (see Fig. 3). If only 12.5% of Co are in the HS state this would give a magnetostriction for the whole crystal of $\varepsilon_v^{(ii)} \approx 0.024/8=0.003$. This is in accordance with the observed volume strain.

We conclude, therefore, that only the scenario (ii) yields a consistent description of our experimental data. In addition, the appearance of a small distortion as a secondary effect can be understood by the fact that the HS states shown in Fig. 4 are also Jahn-Teller active. The positive tetragonal strain $\varepsilon_d$ which is observed in our experiment is consistent with an out-of-plane magnetization. Note that a large distortion is expected for model (i), because the distortion is needed to stabilize the IS state (the IS state cannot be stabilized by a pure volume expansion). In model (ii) the distortion can be a second order effect, a volume expansion is sufficient to stabilize the HS state. A distortion may aid this, but is not essential and can be very small (depending on the corresponding elastic constants).

**Conclusion**

Summarizing, we have demonstrated that the measurement of transversal magnetostriction is possible using the optical FBG method, and we show that our data on $LaCoO_3$ can be interpreted with the help of full multiplet configuration interaction cluster calculations. Within this framework our results provide experimental evidence for a mixed HS-LS state scenario in the high field phase of $LaCoO_3$ which confirms a number of interesting investigations and helps to settle a debate, which has been going on for several decades.

In this state, the HS and LS Co sites are correlated and form some sort of short-range or long-range spin-state order. Because of the significant coupling to the lattice the spin crossover at high field and low temperature is associated with a first order phase transition.

**Methods**

**Sample preparation and characterization.** Large single crystals of the rhombohedrally-distorted perovskite cobaltite $LaCoO_3$ were grown by the floating-zone technique[16]. The chemical composition of the crystals was checked by electron probe microanalysis and thermogravimetric analysis, and structural and magnetic characterisation was performed by X-ray diffraction and superconducting quantum interference device (SQUID) magnetometry. The sample was found to be homogeneous and free from traces of foreign phases. A single crystal of dimensions approximately LaCoO$_3$ with 3x3x3 mm$^3$ was used for the magnetostriction measurements. The cube was cut along the quasi-cubic [100] axes.

**The magnetostriction measurements.** The optical Fibre Bragg grating (FBG) method[11] was used in pulsed magnetic fields up to 70 T, and the magnetostriction was measured with the magnetic field along [100] axes of the crystal. The duration of the pulse is about 150 ms. The contact between fibre and sample, important for the reliability of the results, was realized by superglue. Because of the limited space and the minimal bending radius of the fibre, we used a special guide groove to bend the fibre smoothly, which made it possible to perform transversal measurements in a 10 mm free sample space.

**Acknowledgements**

Part of this work has been supported by EuroMagNET II under the EC contract 228043. We acknowledge the support of the HLD at HZDR, member of the European Magnetic Field Laboratory (EMFL).

**Author contributions** M. Rotter developed the research strategy, performed the electronic structure calculations, and drafted the manuscript. Z. -S. Wang, M. Doerr and M. Rotter did the magnetostriction experiments, D. Prabhakaran prepared the samples, A. Tanaka contributed the XTLS software, A. T. Boothroyd and A. Tanaka contributed essentially to the interpretation of the data. All authors contributed to the final manuscript.

**Additional information**

**Competing financial interests:** The authors declare no competing financial interests.

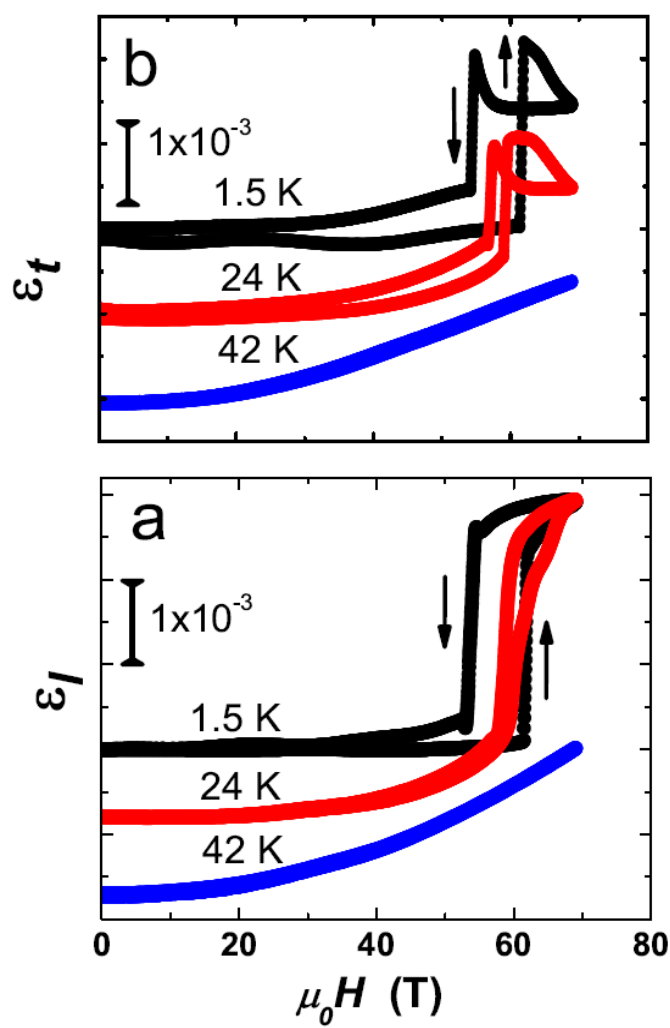

**Figure 1 | Magnetostriction of single-crystalline LaCoO$_3$ measured in pulsed magnetic fields.**
(a) Longitudinal and (b) transversal magnetostriction along the *a*-direction of the crystal.

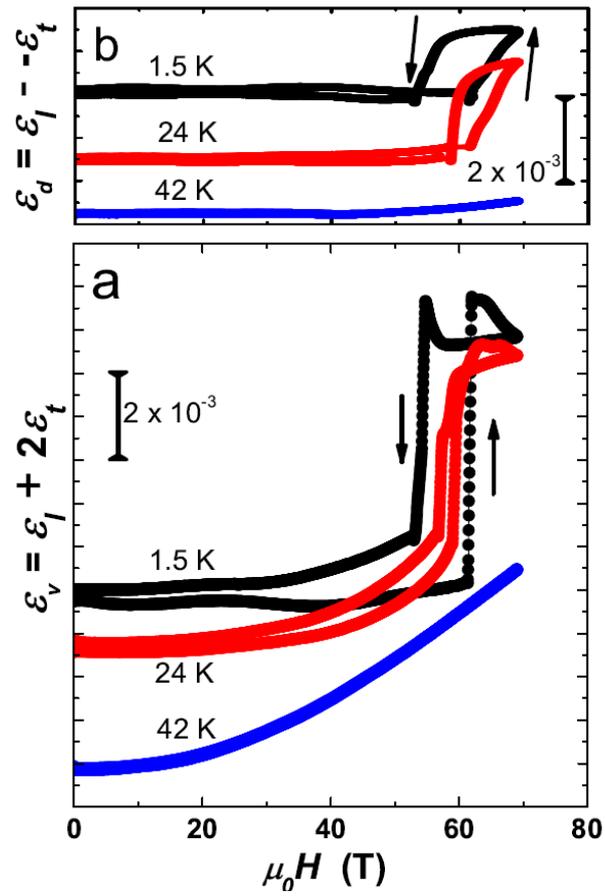

**Figure 2 | Volume effect and tetragonal distortion of LaCoO$_3$ calculated from magnetostriction.**
(a) Volume effect and (b) tetragonal distortion calculated from the longitudinal and transversal magnetostriction as indicated in the vertical axis lables.

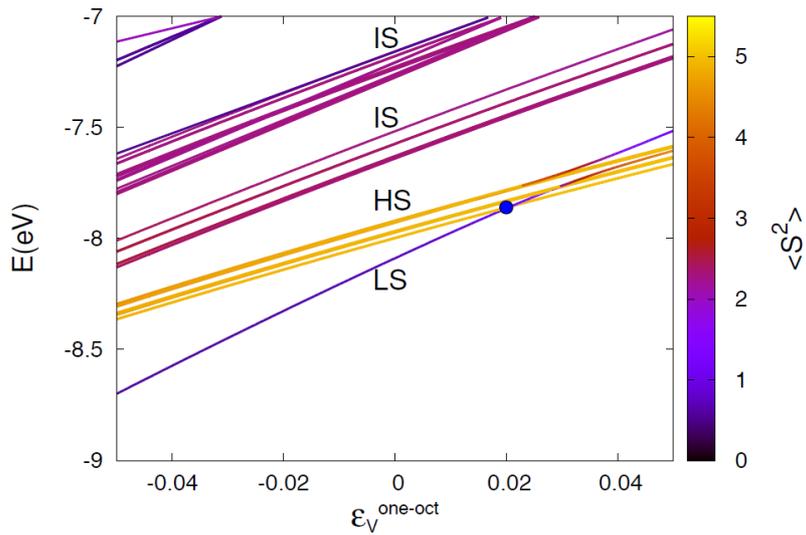

**Figure 3 | Energy levels of a distorted Co-O$_6$ octahedron in LaCoO$_3$ for a symmetry-conserving oxygen displacement.** At zero distortion the ground state is a LS state. The excited HS and IS states are also indicated. Colors correspond to $\langle S^2 \rangle$, the mean squared spin. The transition points discussed in the text are marked by blue dots. Small level splittings ∼ 0.1 eV within the HS and IS levels are caused by the 3$d$ spin-orbit interaction.

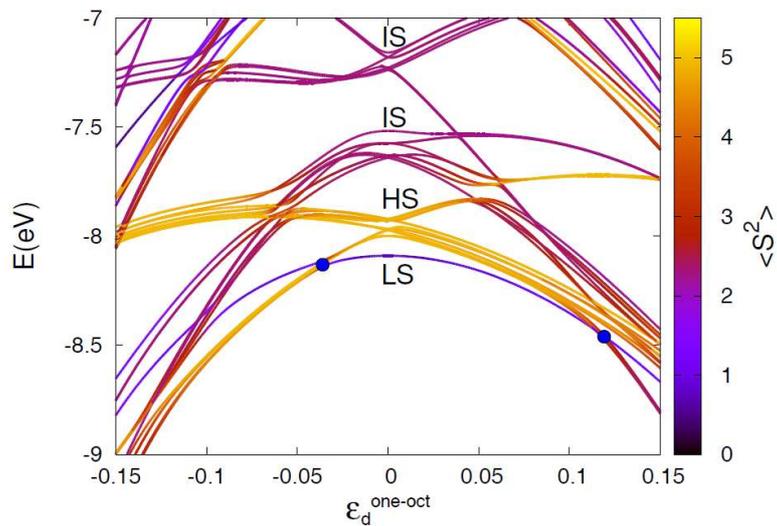

**Figure 4 | Energy levels of a distorted Co-O$_6$ octahedron in LaCoO$_3$ for a volume-conserving tetragonal oxygen displacement.** Colors correspond to $\langle S^2 \rangle$. The transition points discussed in the text are marked by blue dots.